\documentclass[twocolumns]{M-PSI}

\usepackage{array}
\usepackage{amssymb} %
\usepackage{graphicx} %
\usepackage{float} %
\usepackage{amsmath}
\usepackage{pifont}
\usepackage{multirow}
\usepackage{tabularx}

\definecolor{mygray}{gray}{0.2}
\newcommand{\mycross}{\scalebox{0.8}{\textcolor{mygray}{\ding{55}}}}
\title{Listen to the Features: Voice Anonymization Driven by Content Embedding Matching over Signal Reconstruction}

\author{
\AuthorEntry{Adrien~Schneider}{,}{1}{https://adrien-schneider.github.io/}{0009-0005-4874-3940}
\AuthorEntry{Kacper~Zabkowski}{,}{1}{}{}
\AuthorEntry{Anderson~Augusma}{,}{1}{https://augusmaa.github.io/}{0009-0005-0634-0318}
\AuthorEntry{Frédérique~Letué}{,}{2}{https://membres-ljk.imag.fr/Frederique.Letue/}{0009-0005-0634-0318}
\AuthorEntry{Maria~Camila~Pinzon}{,}{1}{}{}
\AuthorEntry{Dominique Vaufreydaz}{}{1}{https://research.vaufreydaz.org/}{0000-0002-8825-0973}\\
\vspace{0.5em}
\textnormal{\normalsize{
{$^1$ Univ. Grenoble Alpes, CNRS, Grenoble INP, LIG, 38000 Grenoble, France}\\ %
{$^2$ Univ. Grenoble Alpes, CNRS, Grenoble INP, LJK, 38000 Grenoble, France}\\
}}
\vspace{0.5cm}
} %


\usepackage{comment}

\begin{document}

\begin{abstract}

The paper presents a voice anonymization model focusing on preserving content rather than producing realistic speech.
It relies on content embeddings extracted from a frozen pretrained wav2vec2 encoder.
These embeddings are decoded into an anonymized signal using vector quantization and a HiFi-GAN vocoder, both trained on LibriTTS without any waveform reconstruction loss or speaker embedding mapping.
The training objective encourages that embeddings of the anonymized signal match those of the original speech.
Results show that this straightforward approach achieves a very low WER, while removing sufficient speaker-specific information to qualify for the VPC's first evaluation condition. 
Emotional content is also partially preserved.
An adversarial auxiliary speaker classifier is then added during training to improve the anonymization performance, but the results show that this fails to improve the privacy–utility trade-off in the current experimental setup.

\end{abstract}

\section{Introduction}

Reproducible research is important for accelerating scientific progress either in Artificial Intelligence, in Social Science research or in Computational Social Science.
In these contexts, sharing data is as important as sharing source code but it becomes increasingly complex due to legal constraints, such as the GDPR \cite{eu_gdpr_2016} and the AI Act~\cite{eu_ai_act_2024} in Europe and other regulations around the world, or even simply considering the need to address ethical considerations.
These rules are beyond discussion, as protecting the privacy of individuals is critical in the current numerical world but one must consider their impact on science discovery due to the limitations they impose on research data sharing.

Among standard usages of privacy algorithms, anonymization of audio and video data is one way to leverage data sharing.
Numerous high-value scientific datasets containing identifiable voices and faces are collected by researchers worldwide.
If these data are sufficiently anonymized, they can be shared to the research community and thus can benefit many other researchers.
To achieve this, anonymization must preserve the semantic content of speech, emotions, interpersonal interactions, and other social signals expressed in recorded videos, thus maintaining the value of the collected data.
A long-term objective for privacy-safe reproducible research would be to make available lightweight anonymization models capable of anonymizing voice directly while recording speech for corpus collection.

This research on voice anonymization is part of a broader project about social-aware video anonymization targeting both audio and video signals.
In this first proposal, this research focuses on voice anonymization preserving what is said while evaluating preservation of voice emotions.
The originality lies in the ability of the model to learn to generate an anonymized audio segment that, once processed by a given speech encoder, would give the same latent content representation than the original voice segment.
There are no constraints nor losses within the training process to force the generation of a realistic, human-sounding speech signal.
The motivation is that, with automated downstream tasks in mind, such as Automatic Speech Recognition (ASR) or Speech Emotion Recognition (SER), the anonymization model only needs to generate any signal that carries meaningful information, without expecting it to be clean speech.
The proposed method also has the advantage of not using pseudo-speakers to do voice conversion-based voice anonymization. As such, it cannot be used to create deepfakes or impersonate the voice of another person without their consent.

The proposed model is evaluated according to the official Voice Privacy Challenge 2026 protocol~\cite{miao2026voiceprivacy}, and was submitted to Track 1 of the challenge, which focuses exclusively on English speech.

The title of this article starts with ``\textit{Listen to the Features}'' as the goal is to develop a straightforward, lightweight model based solely on the features from a pre-trained foundation model.
The aim is to answer the research question: is it possible to ``\textit{listen to the features}'', i.e. to use them without any or with minimal anonymization supervision to achieve good speech recognition performance on an anonymized voice.

Remaining of this paper presents the following contributions:
\begin{itemize}
    \item A lightweight neural architecture driven by content embedding for voice anonymization targeting good WER performance. As previously stated, the proposed model does not use a waveform reconstruction loss, nor speaker embeddings. Inference model is lightweight with less than 99M parameters comprising 18M of trainable ones.
    \item Experiments showing that the best hyperparameter set performs very well on ASR (word error rate of 2.53\%), fair on SER (unweighted average recall of 43.91\%) while being on the lowest privacy-preservation challenge class (equal error rate between 10 to 20\%).
    \item Qualitative automated Mean Opinion Score (MOS) analysis of anonymized voice, highlighting that even if quality of anonymized signals is clearly degraded (MOS of 1.63), the generated signals remain usable for automatic speech recognition, and to some extent for emotion classification. This observation raises a relevant question regarding the required perceptual quality of anonymized signals as a function of the targeted downstream tasks.
\end{itemize}

\section{Related work}

Voice anonymization aims to transform a voice signal in such a way as to conceal the identity of the original speaker whilst preserving the information required for downstream speech processing tasks.
This objective has been formalised as part of the Voice Privacy Challenge series, in which anonymization systems are evaluated according to criteria relating to both privacy and utility~\cite{tomashenko2022voiceprivacy, tomashenko2026third,miao2026voiceprivacy, tomashenko2024voiceprivacy}.
The Voice Privacy Challenge 2026 edition further emphasizes stronger attacker models and multilingual anonymization, making the preservation of linguistic and paralinguistic information particularly important~\cite{miao2026voiceprivacy}.

Several families of voice anonymization methods have been explored in previous research.
Signal-processing approaches, such as transformations based on the McAdams coefficient, modify the spectral envelope of speech in order to reduce speaker identity while keeping the signal relatively intelligible~\cite{patino2020speaker}.
The B2 baseline of the Voice Privacy Challenge 2026 uses such an approach~\cite{miao2026voiceprivacy}.
Other systems rely on speaker embedding replacement, often using x-vectors~\cite{snyder2018x} or similar speaker representations, where the original speaker embedding is substituted or modified before speech resynthesis~\cite{tomashenko2024voiceprivacy,leang2025phd}.
Self-supervised speech representations have become increasingly important for speaker anonymization because they can encode robust linguistic information without requiring explicit phonetic annotations. \texttt{wav2vec2} learns speech representations from raw audio using self-supervised pretraining and has shown strong performance for ASR after fine-tuning~\cite{baevski2020wav2vec}. Such representations have been used in voice anonymization systems to retain linguistic information while suppressing speaker-specific features, such as in baseline B5 of the challenge~\cite{miao2026voiceprivacy}.
Vector-quantized representations provide another mechanism for controlling the amount and type of information retained by an anonymization system.
VQ-VAE introduced discrete latent representations learned through a codebook, creating a bottleneck between the encoder and decoder~\cite{van2017neural}. 
In voice anonymization, such bottlenecks can be used to force the model to preserve task-relevant content while discarding fine-grained speaker-specific details~\cite{leang2024exploring}.
The B5 baseline uses a vector-quantized acoustic bottleneck to extract linguistic representations for anonymized speech resynthesis~\cite{miao2026voiceprivacy}.
Neural vocoders are commonly used to reconstruct waveform audio from intermediate acoustic representations. HiFi-GAN is a generative adversarial vocoder designed for efficient and high-fidelity waveform synthesis~\cite{kong2020hifi}.
It has been used in several Voice Privacy anonymization pipelines~\cite{tomashenko2024voiceprivacy, miao2026voiceprivacy} to generate an output waveform from modified or compressed representations, rather than directly manipulating the original waveform.
In parallel, adversarial learning can be used to reduce the amount of speaker information encoded in the learned representation.
Gradient reversal, originally introduced for domain-adversarial learning, enables a model to learn representations that are useful for the main task while being uninformative for an auxiliary classifier~\cite{ganin2015unsupervised}.
This principle has been applied to voice anonymization to suppress speaker-identifying information from learned representations~\cite{perero2022adversarial}.
An auxiliary speaker predictor can therefore be used to encourage the learned representation to discard speaker-specific characteristics.

The model proposed in this article is designed to be straightforward and lightweight, with a particular emphasis on preserving linguistic content for downstream ASR.
Rather than optimizing a conventional waveform-reconstruction objective, the model is trained to match the \texttt{wav2vec2}-based content representations of the original and anonymized speech.
Furthermore, unlike speaker-embedding-based approaches, the proposed model does not require any explicit pseudo-speaker representation for waveform reconstruction.
Instead, it relies on an auxiliary speaker-prediction objective to encourage the model to discard speaker information while preserving the linguistic content of the input speech.

\section{System description}

\subsection{Architecture} \label{archi}

\begin{figure*}[ht]
  \centering
  \includegraphics[width=\linewidth]{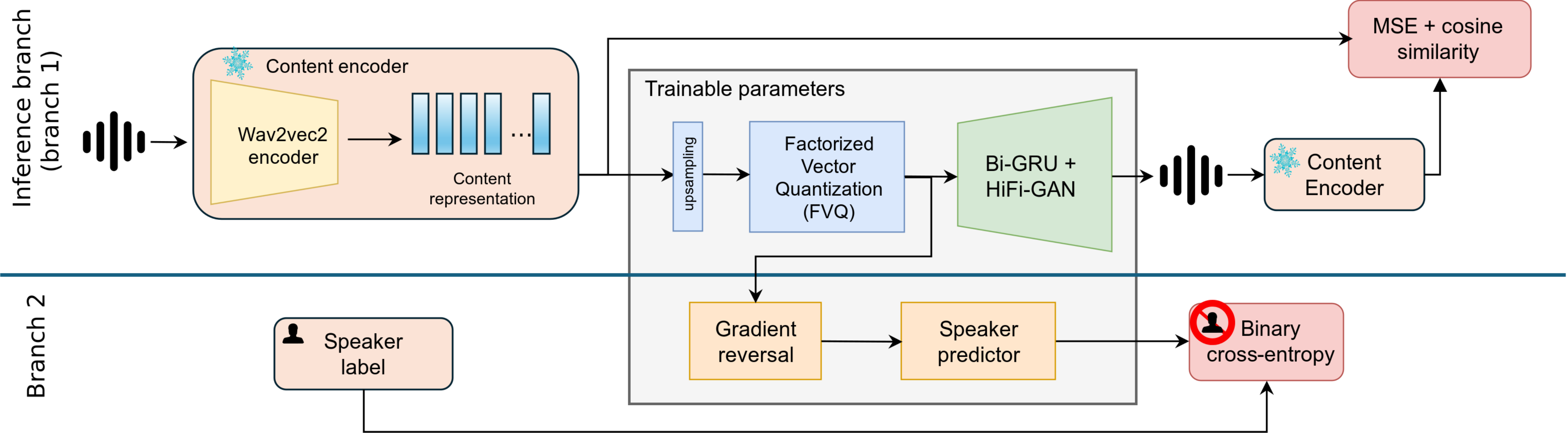}
  \caption{Schematic diagram of model. The network architecture has 2 branches. The first one starts with a frozen \texttt{wav2vec2} encoder. Then embeddings transit thought a Factorized Vector Quantization followed by a bi-GRU/HiFi-GAN process to generate the anonymized waveform. This waveform output is passed again into the frozen content encoder to generate output content embedding. MSE and cosine similarity losses are used between input and output content embeddings to train this branch. The second network branch enforces the system to be unaware of the current speaker thanks to gradient reversal, a speaker predictor network and a cross-entropy between input and predicted speaker labels.}
  \label{fig:system_training}
\end{figure*}

During training, as depicted on figure~\ref{fig:system_training}, the model uses a two-branch approach with different functions.
The first branch uses a pre-trained encoder meant to extract latent content representations from a speech signal.
Embeddings transit into a quantization module followed by a waveform generator to reconstruct an anonymized speech from quantized latent representation.
This branch corresponds to the final inference model.
The second branch is used only at training time.
It inflects training to optimize network weights to avoid keeping speaker information.
These modules are described in the remainder of this section.

The first branch of the model uses a frozen pretrained \texttt{wav2vec2}~\cite{baevski2020wav2vec} model for speech encoding.
The HuggingFace implementation is used, specifically the \texttt{wav2vec2-base-960h} model\footnote{\url{https://huggingface.co/facebook/wav2vec2-base-960h}}.
This is the standard version of the model, pretrained and fine-tuned on 960 hours of speech from the LibriSpeech dataset~\cite{panayotov2015librispeech}.
The latent representations are extracted from the ninth hidden layer of the model.
The content embeddings coming from the \texttt{wav2vec2} encoder go through a trainable upsampling layer, designed to match the operating needs of the HiFiGAN vocoder (160 samples per frame vs. roughly 320 output by \texttt{wav2vec2}~\cite{baevski2020wav2vec}).
This layer is a 1-dimensional transposed convolution.
The upsampled content embeddings are then quantized by the Factorized Vector Quantization (FVQ) module 
inspired by the quantizer-vocoder combination for voice conversion proposed in \cite{leang2025phd}. 
As it employs a single codebook, the proposed architecture does not differentiate between linguistic content and other features such as emotional content as the referenced model did.
It also differs by codebook size and codeword length.
The core idea is to create a bottleneck forcing the model to retain features that are useful for the downstream task and to discard the rest.
In section \ref{sec:experiments}, different values of codebook size and codeword length, and their impact on challenge metrics are investigated.
The quantized content representations are then processed by a Bi-directional Gated Recurrent Unit (bi-GRU) before being fed to a HifiGAN vocoder, which generates an acoustic signal of approximately the same length as the input signal.

A second branch is added to the network architecture at training time to prevent the model from retaining information about the specific voice of the original speaker.
Derived from the adversarial classifiers proposed in \cite{naturalspeech3}, this branch consists of an auxiliary speaker predictor combined with a gradient reversal module. The speaker classifier consists of successive weight-normalized 1D-convolutional layers followed by Snake-Beta activation functions \cite{lee2023bigvgan, liu2020snake}.
The speaker predictor takes the quantized latent representations produced by the FVQ as input and outputs a speaker id, i.e. a unique number for each speaker.

\subsection{Training process}

During the training phase, all the weights of both network branches are optimized except for the frozen \texttt{wav2vec2} encoder.
Branch divergence starts after the FVQ output.
From these projected content embeddings, the first branch uses the Hifi-GAN generator fully retrained to generate the output audio signal.
The discriminators of the original Hifi-GAN are not used.
The generated signal is then fed to the frozen \texttt{wav2vec2} encoder to get its content embeddings.
The objective loss of the model targets producing output-signal embeddings that match the original ones as closely as possible.
Inspired by the speaker embedding generation loss used by Liu at al. in IDmap \cite{idmap}, the difference between generated and original embeddings is measured using a content loss function $L_c$, which is an equally weighted sum of Mean Squared Error (MSE) and cosine similarity losses between the original and generated embeddings.
The loss of the first branch is computed as
\begin{equation}
    L_c = 0.5\cdot|x-y|_2^2+0.5\cdot\left(1-\frac{x^\top y}{|x|_2\,|y|_2}\right)
    \label{eq:content_loss}
\end{equation}
where $x$ and $y$ are respectively embeddings of input and generated audio signals.

For the second branch, a cross-entropy loss $L_s$ is used to compare the output of the auxiliary speaker predictor with the ground-truth speaker labels.
The total loss is the weighted sum of $L_c$ and $L_s$ as
\begin{equation}
    L = \lambda_s L_s + \lambda_c L_c    
\end{equation}
where $\lambda_s$ and $\lambda_c$ are weight factors for the speaker and content losses respectively.

\section{Experiments and results}
\label{sec:experiments}

\subsection{Evaluation protocol}

The official Voice Privacy Challenge 2026 evaluation protocol~\cite{miao2026voiceprivacy} simulates a game between a \textit{user} and an \textit{attacker}. 
The user shares anonymized utterances, referred to as \textit{trial} utterances, for downstream ASR and SER tasks. 
The attacker's objective is to infer the identity of the user from those anonymized utterances. 
The attacker is assumed to be \textit{semi-informed}, meaning that they have access to the anonymization system employed by the user, as well as to several anonymized utterances from each user, referred to as \textit{enrollment} utterances. 
This enables the attacker to train an Automatic Speaker Verification (ASV) system specifically adapted to the anonymization system under evaluation. 
Privacy preservation is assessed using a \textit{privacy} metric, while the preservation of linguistic and emotional content is evaluated using two \textit{utility} metrics based on ASR and SER tasks. 

The proposed model is evaluated using the tools provided by the Voice Privacy Challenge 2026 organizers.
As explained above, in the semi-informed scenario, the attacker has access to the anonymization model being evaluated and uses it to anonymize the enrollment utterances in order to reduce the mismatch with the trial utterances. 
The attacker then anonymizes the \textit{LibriSpeech-train-clean-360} dataset using the same model and trains an Automatic Speaker Verification (ASV) system on the resulting data. 

This system is then used to compute the privacy metric, the Equal Error Rate (EER). 
Given an utterance, the ASV system outputs a speaker vector intended to represent the speaker's identity. 
The attacker computes a cosine similarity score between the speaker vectors of the enrollment and trial utterances.
This score is compared with a threshold to determine whether the two utterances are from the same speaker. 
The EER is the error rate at the threshold $\theta$ for which the false-positive rate equals the false-negative rate. 
A higher EER indicates better privacy.

The ability of the anonymization system to preserve useful information is evaluated on ASR and SER tasks using Word Error Rate (WER) and Unweighted Average Recall (UAR), respectively.
The challenge tools use a \texttt{wav2vec2}-based ASR system trained on the original (non-anonymized) \textit{LibriSpeech} to transcribe the anonymized utterances.
The transcriptions of the anonymized utterances are compared with the original ones, and the WER is computed as the sum of substitutions $N_{\mathrm{sub}}$, deletions $N_{\mathrm{del}}$, and insertions $N_{\mathrm{ins}}$ divided by the number of words in the original transcription $N_{\mathrm{ref}}$:
\begin{equation}
    \mathrm{WER} = \frac{N_{\mathrm{sub}} + N_{\mathrm{del}} + N_{\mathrm{ins}}}{N_{\mathrm{ref}}}
\end{equation}
A lower WER therefore indicates higher utility for ASR.

The challenge's SER evaluation system is also based on \texttt{wav2vec2}. It classifies utterances into $N_{\mathrm{class}}$ emotion classes, from which the UAR is computed as 
\begin{equation}
    \mathrm{UAR} = \frac{\sum_{i=1}^{N_{\mathrm{class}}} R_i}{N_{\mathrm{class}}}
\end{equation}
where the recall $R_i$ for class $i$ is defined as the number of true positives  divided by the total number of samples in that class. A higher UAR indicates higher utility for SER.

The EER and WER are computed on the anonymized development and evaluation subsets of \textit{LibriSpeech}, while the UAR is computed on the anonymized development and evaluation subsets of \textit{IEMOCAP}.

\subsection{Hyperparameter exploration}

Several experiments were conducted to identify the best combination of hyperparameters, notably codebook size and loss weights. 
All experiments described in this section were conducted by training the models on NVidia A100 GPUs.
The AdamW optimizer was used, with betas of 0.8 and 0.99.
An exponential learning rate decay of 0.999 was applied.
For training, audio data was resampled to 16 kHz and split into random 16640-samples-long segments, which gives 52 frames of audio data per sample.
The number of parameters of the model is summarized in table \ref{tab:model_params}.
Training was done on LibriTTS \cite{libritts} using train-clean-100 and train-clean-360 subsets.

\begin{table}[hb]
\centering
\caption{Number of model parameters (in millions). For the encoder, only parameters until hidden layer 9 are included. The speaker predictor is not used at inference time, hence the additional total in the last line.}
\label{tab:model_params}
\begin{tabular}{lcc}
\toprule
\textbf{Component} & \textbf{Total (M)} & \textbf{Trainable (M)} \\
\midrule
Generator & \hphantom{0}18.54 & \hphantom{0}18.54 \\
\hphantom{000}Upsampling + FVQ & \hphantom{00}2.38 & \hphantom{00}2.38 \\
\hphantom{000}BiGRU + HifiGAN & \hphantom{0}16.17 & \hphantom{0}16.17 \\
Speaker predictor & \hphantom{0}14.21 & \hphantom{0}14.21 \\
Encoder   & \hphantom{0}80.20 & \hphantom{00}0.00  \\
\midrule
Total      & 112.94 & \hphantom{0}32.75 \\
\midrule
Total (inference model)      & \hphantom{0}98.73 & \hphantom{0}-- \\
\bottomrule
\end{tabular}
\end{table}

The detailed configurations are listed in table \ref{tab:configurations}, along with their respective score on the challenge metrics.
Some evaluation results were not available at the time of submission because of errors occurring during ASV inference. Those errors could not be investigated on time. Those missing results are marked with dashes in the aforementioned table.

\subsubsection{Content-only experiments}

The first reported results are from experiments where the speaker loss was set to 0, i.e. only the content information was taken into consideration.
The goals are to (i) verify that the model is actually able to generate a signal that conveys what the speech encoder identifies as content, and (ii) get a first idea of the impact of the size of the codebook.
Two different codebook sizes are evaluated: 2048 (with a codeword length of 8) and 65536 (with a codeword length of 16).
The first configuration is intended to test a strong bottleneck, while the latter is meant to match the combined codebook sizes of \cite{leang2025phd}.
Their respective experiment codes in table \ref{tab:configurations} are \texttt{cb2048} and \texttt{cb65536}.
They were both trained with a batch size of 128 for about 50 epochs.

Results show that the target performance on downstream task, i.e. automatic speech recognition, is reached.
The \texttt{cb65536} model (65536 codewords of length 16) reaches a WER of 2.53, which is very close to the original data and better than the challenge baseline systems.
The second model with fewer codewords had a WER of around 11\%.
One interesting result is that \texttt{cb65536} reaches 43.91 UAR without any supervision on emotion (around 27 pp below original data).
It confirms that \texttt{wav2vec2} embeddings also encode emotional information.
Without any supervision for anonymization, the EER for both these experiments falls into the lower considered rank for the challenge (between 10 and 20\%).

\subsubsection{Experiments including the speaker predictor}

The experiments with a code consisting of a single lowercase letter from \textit{a} to \textit{e} in table \ref{tab:configurations} represent experiments where the speaker loss is not null.
They have been designed to see what kind of impact light speaker anonymization supervision has on WER performance when using foundation embeddings.
In all these runs, the contribution of the speaker loss to the total loss is lower than 10\%, more than 90\% remaining for the content loss.
They were all conducted for 100 epochs, with a batch size of 128. The evaluated content loss to speaker loss weight ratio were 10/1, 20/1 and 40/1.

In the gathered results, one can see in table~\ref{tab:configurations} that the anonymization supervision has a strong impact on the proposed model.
Indeed, WER increases up to 100 except for the \texttt{e} configuration (same number of codewords as \texttt{cb65536}, former best model) with 60\% of WER using a 40/1 ratios between respectively content and speaker losses.
In this last case, even if speaker loss contributes to less than 4\% to total loss, it degrades drastically WER performance.
The same reasoning applies to UAR results.
Adding the anonymization supervision degrades UAR performance but in a smaller margin that WER,
the best system (\texttt{e}) remaining at 38.68 UAR.

\subsubsection{Results with updated evaluation tools}

After submission, the organizers have reproduced the evaluations with an updated version of the evaluation tools.
The update fixes a bug in the speaker embedding normalization and updated the trial list for privacy evaluation for Track 1.
As a result, EER values for Track 1 decrease by approximately 2.48 percentage points on average.
The final rankings are based on these updated results.
The updated values for our \texttt{cb65536} submission are presented in Table \ref{tab:updated-results}, along with the baselines.

\begin{table*}[t]
    \centering
    \caption{Results of the proposed model in its best configuration (\texttt{cb65536}) compared to original data and challenge baselines. 
    All metrics were computed using the tools provided by the challenge organizers. For SER utility, the UAR is the average over all emotions. 
    Numbers in bold font indicate the best value of the column regarding the challenge goal. 
    Only the metrics computed on the test subsets are used for ranking.}
    \resizebox{\linewidth}{!}{
    \begin{tabular}{l|cc|cc|cc}
        \toprule
        \multirow{2}{*}{\textbf{System}} & \multicolumn{2}{c|}{\textbf{Privacy -- EER} (\%, semi-informed) $\uparrow$} & \multicolumn{2}{c|}{\textbf{ASR utility -- WER} (\%) $\downarrow$} & \multicolumn{2}{c}{\textbf{SER utility -- UAR} (\%) $\uparrow$} \\
               & LibriSpeech-dev & LibriSpeech-test & LibriSpeech-dev & LibriSpeech-test & IEAMOCAP-dev & IEMOCAP-test \\
        \midrule
        Orig.            & \hphantom{0}7.34 & \hphantom{0}3.91 &          \hphantom{0}1.80 &          1.84 &          69.08 &         71.06 \\
        \midrule
        B2               & \hphantom{0}6.97 & \hphantom{0}4.71 &                     10.44 &          9.96 & \textbf{55.66} & \textbf{53.49} \\
        B3               &            19.14 &            16.85 &          \hphantom{0}4.31 &          4.31 &          38.08 &         35.24 \\
        B4               &            17.96 &            14.33 &          \hphantom{0}6.16 &          5.90 &          41.97 &         42.78 \\
        B5               &   \textbf{24.20} &   \textbf{21.28} &          \hphantom{0}4.91 &          4.44 &          40.08 &         38.25 \\
        \midrule
        \texttt{cb65536} &            15.36 &            11.43 & \textbf{\hphantom{0}2.47} & \textbf{2.60} &          44.65 &         43.17 \\
        \bottomrule
    \end{tabular}
    }  %
    \label{tab:results}
\end{table*}

\begin{table}[t]
  \centering
  \caption{Results on the test subsets reproduced by the organizers with an updated version of the evaluation tools. Those results will be used for the final ranking. The EER values correspond to the semi-informed attacker scenario.}
    \begin{tabular}{lrrr}
      \toprule
        \textbf{System}  & \textbf{EER $\uparrow$} & \textbf{WER $\downarrow$} & \textbf{UAR $\uparrow$} \\
        \midrule
        Orig.            &          3.41  &         1.84  &         71.08  \\ \midrule
        B2               &          3.72  &         9.11  & \textbf{53.52} \\
        B3               &         14.24  &         4.26  &         37.70  \\
        B4               &         11.13  &         5.90  &         42.78  \\
        B5               & \textbf{17.05} &         4.53  &         38.09  \\ \midrule
        \texttt{cb65536} &          8.92  & \textbf{2.59} &         43.20  \\
      \bottomrule
    \end{tabular}
  \label{tab:updated-results}
\end{table}

\begin{table*}[t]
\centering
\caption{%
Summary of the experiment configurations.
The first column represents the configuration names used for the official VPC submissions\protect\textsuperscript{1}.
The next two columns show the codebook size and codeword length that were used in each configuration.
The next three columns contain the weights applied to the content and speaker losses. 
The contribution of the content loss to the final loss is also given as a percentage.
The final three columns give the challenge evaluation results on the datasets used for ranking (IEMOCAP-test for SER, LibriSpeech-test for the rest).
Although the EER results in the lazy-informed scenario are not used for ranking, they are included for comparision of the configurations, 
since the semi-informed EER computation failed in some cases\protect\textsuperscript{2}. The missing EER results are indicated by dashes.}
\label{tab:configurations}
\resizebox{\linewidth}{!}{

\begin{tabular}{l|cc|ccc|cccc}
\toprule
\textbf{Config\textsuperscript{1}} &
\textbf{Codebook} &
\textbf{Codeword} &
\textbf{Content} &
\textbf{Speaker loss} &
\textbf{$\lambda_c$ contrib.} &
\multicolumn{2}{c}{\textbf{EER $\uparrow$ (\%)}} &
\textbf{WER $\downarrow$} &
\textbf{UAR $\uparrow$} \\
\textbf{name}&
\textbf{size}&
\textbf{length}&
\textbf{loss} ($\lambda_c$)&
\textbf{weight} ($\lambda_s$)&
\textbf{to loss} (\%)&
lazy-informed &
semi-informed &
(\%) &
(\%) \\
\midrule
\texttt{cb2048}  & \hphantom{0}2048  & \hphantom{0}8  & \hphantom{0}1 & \mycross & 100.00 & 22.17 & 12.03 & \hphantom{0}11.37 & 39.43 \\
\texttt{cb65536} & 65536 & 16 & \hphantom{0}1 & \mycross & 100.00 & 21.16 & 11.43 & \hphantom{00}\textbf{2.60} & \textbf{43.17} \\
\midrule
\texttt{a} & \hphantom{0}2048  & \hphantom{0}8  & 10 & 1 & \hphantom{0}90.91 & 33.50 & \textbf{19.48} & 105.49 & 32.30 \\
\texttt{b} & \hphantom{0}2048  & \hphantom{0}8  & 20 & 1 & \hphantom{0}95.24 & 33.70 & --     & 124.07 & 30.45 \\
\texttt{c} & 65536 & 16 & 10 & 1 & \hphantom{0}90.91 & \textbf{46.22} & --     & \hphantom{0}99.99 & 25.63 \\
\texttt{d} & 65536 & 16 & 20 & 1 & \hphantom{0}95.24 & 44.13 & --     & 100.00 & 23.44 \\
\texttt{e} & 65536 & 16 & 40 & 1 & \hphantom{0}97.56 & 26.58 & 13.67     & \hphantom{0}62.53 & 37.04 \\
\bottomrule
\end{tabular}
}\\ %
\begin{minipage}{\linewidth}
\footnotesize{\textsuperscript{1}~The prefix \texttt{\_mpsi\_} can be added to the configuration codes to retrieve the anonymization suffixes used for results submission to the challenge.}\\
\footnotesize{\textsuperscript{2}~The evaluation tools converge to \textit{NaN} values failing training of the attacker.}
\end{minipage}
\end{table*}

\subsection{MOS evaluation}

\begin{table}[ht]
  \centering
  \caption{Automated Mean Opinion Score (MOS) using 3 automated MOS annotators. Values range from 1 (bad quality) to 5 (high quality). Values are reported for original and anonymized \textit{LibriSpeech} train-clean-360 using the \texttt{cb65536} model.}
  \begin{minipage}{\linewidth}
    \footnotesize %
    \begin{tabularx}{\linewidth}{l|ccc|ccc}
    \multicolumn{1}{c}{\multirow{2}[0]{*}{Model}} & \multicolumn{3}{c}{Original \textit{LibriSpeech}} & \multicolumn{3}{c}{Anonymized \textit{LibriSpeech}} \\
         & min   & avg  & max   & min   & avg  & max \\
    \toprule
    UTMOS~\cite{saeki22c_interspeech}  & 1.38     & 4.10     & 4.57     & 1.29     & 1.80     & 2.67 \\
    SHEET \cite{sheet}                 & 2.02     & 4.22     & 4.61     & 1.24     & 1.35     & 2.29 \\
    RAMP+ \cite{wang2025ramp+}         & 1.56     & 4.00     & 4.45     & 1.11     & 1.75     & 2.51 \\
    \midrule
    Mean                               & 1.65     & 4.11     & 4.54     & 1.21     & 1.63     & 2.49 \\
    \bottomrule
    \end{tabularx}%
  \end{minipage}
  \label{tab:MOS experiment}%
\end{table}%

The strong voluntary bias of this research is to focus training objectives on downstream tasks, mainly ASR, and to avoid using reconstruction or signal quality losses.
However, it is still interesting to see to what extent the reconstructed signal happen to look like natural speech. Informal listening tests highlighted that the anonymized utterances sounded ''\textit{robotic}``, but most of the words could still be recognized by a human ear.
ASR evaluation system (see table~\ref{tab:configurations}) are also able to deal with the anonymized signals.
One needs to quantify the former perceptive results and to provide a more complete qualitative evaluation. An evaluation of the Mean Opinion Score (MOS) is provided.
As its name suggests, MOS consists of mean quality score given by several humans ranging on a scale from 1 (bad quality) to 5 (high quality).
As conducting a full human MOS evaluation during the challenge time is not feasible, automated MOS estimation is performed.
To comply with MOS formalism, the decision was made to get MOS from three different models and compute their mean score:
UTMOS \cite{saeki22c_interspeech}, SHEET \cite{sheet} and RAMP+ \cite{wang2025ramp+}.

The Results are depicted in table~\ref{tab:MOS experiment}.
MOS is evaluated on both the original \textit{LibriSpeech} train-clean-360 and its anonymized counterpart using the best model from the experiments previously presented (\texttt{cb65536}).
One can see that for original clean dataset, the MOS score is slightly over 4, even if some sentences have bad score (min MOS down to 1.38, 1.65 in average on the three models).
The anonymized counterpart dataset has lower MOS scores.
While min MOS values are not strongly impacted, they are nevertheless getting closer to~1.
In contrast, MOS score lost 2.48 in absolute, down to~1.63. 
The max MOS value is also greatly impacted with~2.49.
These results corroborate the perception of a ``\textit{robotic}'' voice for anonymized speech signals.

\section{Discussion}

The configuration yielding the best results w.r.t. the challenge metrics is \texttt{cb65536}, i.e. a codebook size of 65536, codeword size of 16 and no loss for speaker prediction.
As shown on table \ref{tab:results}, it obtains a WER that is quite close to that on the original data and better than all of the challenge baselines. 
UAR values indicate that emotional information is partially preserved. 
The EER of 11.43 \% places it in the lowest ranking category of the challenge (10\% - 20\%). 
Overall, \texttt{cb65536} outperforms all the baseline systems that would fall in the same ranking category.
Those results suggest that even with no incentive to conceal the speaker's voice, the bottleneck and quantization enable the model to retain most of the useful information -- at least as far as ASR is concerned -- while already discarding speaker-specific information.
Even without supervision, some emotion information is preserved as UAR reaches 43.91, meaning that using foundation models, even those not trained with vocal emotion in mind, is relevant for anonymization.

However, the addition of anonymization supervision using the speaker predictor did not have the expected effect of improving the privacy-utility trade-off, at least with the configurations that have been tested.
As shown on table \ref{tab:configurations}, the models trained with a speaker supervision yield very bad WERs.
This suggests that the loss on speaker prediction also prevents the model from properly capturing content.
Indeed, in listening tests, anonymized utterances were completely unintelligible, except for configuration \texttt{e}, where some words could be recognized.
This hints towards testing other content-to-speaker loss ratios.
Moreover, the speaker predictor used in anonymization supervision is taken from former research~\cite{leang2024exploring, leang2025phd}.
Other speaker anonymization approaches deserve to be explored.
Nevertheless, as we do not want to provide deepfake tools, models using speaker embeddings for anonymization should be avoided.

Another limitation of the present evaluation is the potential for evaluation-model coupling. 
The proposed model is trained to match content representations extracted with \texttt{wav2vec2}, which is also used as the speech encoder in the challenge's evaluation tools. 
Therefore, the reported utility results may partially benefit from a compatibility between the internal representations of the anonymization model and that of the downstream evaluation systems, and do not establish that the proposed model preserves content in a system-independent manner. 
Further experiments with downstream ASR and SER systems based on other families of speech encoder are needed to assess the generalization of the model. 

On the topic of the quality of anonymized signals, the MOS scores factually highlight that the quality is drastically impacted, in line with perception tests made on subsets of the dataset.
Nevertheless, automatic speech recognition and emotion classification are still possible.
In the context of reproducible research, this raises the question of the target quality of the anonymized signals.
Indeed, in a pure machine learning or computational social science approaches, losing hearing quality of the signal is not a problem as far as downstream tasks can be applied successfully.
In social science research where human annotation is still important, anonymization must not degrade the quality of speech.
The trade-off between anonymization and signal quality is therefore important in the context of reproducible research, not only in terms of metric performance but also in terms of the target application.

Last, regarding number of parameters of the model (table~\ref{tab:model_params}), the total size is 112.94M at training time.
Most of the parameters are in the \texttt{wav2vec2} encoder (80.M).
Only 32.75M are trainable parameters including speaker predictor model.
At inference time, the model size is less than 99M parameters, comprising 18.54M trained ones.
Regarding the target automatic speech recognition task and the corresponding WER, size of the model is very competitive and in line with our long-term reproductive research and privacy-preserving goal: be able to record directly anonymized speech while collecting corpora.

\section{Conclusion}

This article introduces a straightforward lightweight voice anonymization model that focuses on preserving linguistic content rather than producing high‑quality human‑like speech signals.
The model operates entirely in the latent space of a frozen \texttt{wav2vec2} encoder and relies on vector quantization and a HiFi‑GAN vocoder to generate the anonymized signal.
No waveform-reconstruction objective nor speaker embedding mapping are used in this proposal.

For the VoicePrivacy 2026 evaluation, the best configuration achieves word error rates close to those obtained on the original data, maintains a fair level of emotion recognition performance and reaches privacy scores that fall within the lowest ranking category of the challenge (10 to 20\%).
These results show that for automated downstream tasks such as ASR and SER, generating a high-quality natural signal is not strictly necessary as long as the anonymized waveform carries a stable and informative content representation.
The strong degradation of perceptual quality, stated by informal listening tests and by automated MOS estimation, confirms that the anonymized voice sounds ``robotic'', yet remaining usable for machine learning.
These results raise again the relevant question for privacy‑preserving reproducible research: the target quality of anonymized signals should be defined in relation to the intended application, whether it is fully automatic processing or human annotation for instance in social science research.
Results also suggest that a simple bottleneck based on factorized vector quantization can discard part of the speaker‑specific information without anonymization supervision.
On the contrary, adding a speaker prediction loss as supervision excessively damages content preservation in the proposed setup.
These findings call for more nuanced training objectives and alternative anonymization strategies that balance privacy, intelligibility and emotional content.
These strategies must not rely on pseudo‑speakers or speaker embeddings to avoid being misused for deepfake generation.
Future research will investigate better trade‑off between anonymization and signal quality, exploring multilingual data and more diverse corpora.

Last, considering this research within the context of models for privacy-preserving data sharing for reproducible research, one must analyze the systems’ usability in terms of computational cost.
While WER, UAR, and EER performance metrics are certainly important, anonymization models must not be resource-intensive, always requiring powerful servers and high-performance GPUs.
This is even more true if one plans to directly record anonymous speech during corpus collection.
Based on this premise, the proposed architecture was designed to use a model with around 99M parameters which is relatively small regarding today’s standards.
Yet, the model performs very well in terms of word error rate but needs to improve in terms of anonymization without drastically increasing the number of its parameters.
There is also a trade-off between metric performance and computational cost that must be carefully questioned in future research, with potential solutions adaptable to different scales.

\section{Acknowledgments}

This research is built upon previous research from Sotheara Leang~\cite{leang2025phd}, a former member of the M-PSI Team and a former participant of the VPC Challenge. This research was supported by the TALISMAN project (ANR-22-CE38-0007).

\section{Generative AI Use Disclosure}
Some authors of the paper are using RAGs as complementary tools for literature review.
Generative AI tools were used for some latex table formatting tweaks and some sentence rephrasing.
AI has been employed to check for orthographic and grammar correctness.

\bibliographystyle{IEEEtran}
\bibliography{biblio}

\end{document}